**Insulating behavior in ultra-thin bismuth selenide field effect transistors**

Sungjae Cho, Nicholas P. Butch, Johnpierre Paglione, and Michael S. Fuhrer

*Center for Nanophysics and Advanced Materials, University of Maryland, College Park, MD 20742-4111, USA*

**Ultrathin (~3 quintuple layer) field-effect transistors (FETs) of topological insulator $Bi_2Se_3$ are prepared by mechanical exfoliation on 300nm $SiO_2$/Si susbtrates. Temperature- and gate-voltage dependent conductance measurements show that ultrathin $Bi_2Se_3$ FETs are *n*-type, and have a clear OFF state at negative gate voltage, with activated temperature-dependent conductance and energy barriers up to 250 meV.**

Topological insulators are new class of materials that have a bulk band gap and gapless Dirac surface states which are topologically protected from back scattering or localization by time-reversal symmetry. The existence of surface states in $Bi_2Se_3$ was observed recently by angle-resolved photoemission spectroscopy (ARPES)[1-3] and scanning tunneling spectroscopy (STS)[4-6]. ARPES measurements indicate $Bi_2Se_3$ has 0.3 eV bulk band gap and a single Dirac cone surface state in the gap, however electronic transport experiments which are dominated by surface state transport remain elusive[7-10], due to bulk conduction in $Bi_2Se_3$.

One strategy to reduce the contribution of bulk conduction is to fabricate very thin $Bi_2Se_3$ layers, and some electronic transport experiments on thin $Bi_2Se_3$ films[11-12] and crystals[13-14] have been reported. An interesting question is: How thin can $Bi_2Se_3$ be while retaining its three dimensional topological insulator character? The thinnest layer that maintains a 2:3 Bi:Se stoichiometry is the quintuple layer (QL) of 5 alternating Se and Bi planes, while the thinnest slab which retains the symmetry of bulk $Bi_2Se_3$ is one unit cell, or three QL units, thick. Theoretical work[15-17] as well as ARPES experiments[18] indicate that in few-QL $Bi_2Se_3$ the surface states may hybridize and open a bulk energy gap, resulting in either a two-dimensional insulator or a quantum spin Hall system with insulating bulk and conducting chiral one-dimensional edge modes[19-21]Here, we report electrical measurements on ultrathin (~1 unit cell, or ~3 QLs) $Bi_2Se_3$ crystals as a function of gate voltage and temperature. We observe clear insulating behavior beyond a threshold gate voltage, with activated energy gaps up to 250 meV. The results indicate that ~3 QL $Bi_2Se_3$ crystals are conventional insulators with energy gaps exceeding 250 meV.

$Bi_2Se_3$ crystals were prepared as described in Reference[7]. Bulk carrier concentrations (*n*-type) were in the range of $2-4 \times 10^{18}$ $cm^{-3}$. $Bi2Se_3$ was mechanically exfoliated on substrate of 300nm $SiO_2$ over a conducting Si back gate using a "Scotch tape" method similar to that used for graphene[22]. Fig.1a shows an optical micrograph of a typical mechanically exfoliated $Bi_2Se_3$ crystal on $SiO_2$/Si. Crystals of thickness 3.5nm-30nm were found, and could be differentiated by color contrast similar to few-layer graphene[23]. The thickness of the samples was measured by atomic force microscopy (AFM), which may overestimate the true thickness of the crystal as is observed for graphene on $SiO_2$[23]. The thinnest samples (thickness $t$ = 3.5nm, corresponding to ~3 QLs) were chosen for this study. Electron beam lithography was used to define Pd electrodes; Fig.1b shows the completed device. No adhesion layer was used; we found that using Cr or Ti as an adhesion layer makes contact resistance increase rapidly with time, which might be related to oxidation of adhesion layer.

Fig. 2 shows the gate-voltage ($V_g$) dependent transport properties of four $Bi_2Se_3$ transistors of various thicknesses. For thicker samples (Sample 1, $t = 14$ nm; and Sample 2, $t = 6.5$ nm), the sheet conductivity measured in a four-probe configuration is shown. For the thinnest samples (Samples 3 and 4, $t = 3.5$ nm) the two-probe conductance is shown as a function of $V_g$. Because of the high sample resistance at low temperatures and negative $V_g$, we were unable to perform four-probe measurements on the thinnest samples. We always observe *n*-type doping in exfoliated $Bi_2Se_3$, and for Samples 1 and 2 the carrier density *n* determined by Hall effect at $V_g = 0$, $n = 1.5 \times 10^{13}$ cm$^{-2}$ for Sample 1, and $n = 2.5 \times 10^{13}$ cm$^{-2}$ for Sample 2, exceeds the density of the surface state at the conduction band edge (~$5 \times 10^{12}$ cm$^{-2}$ for one surface, or $1 \times 10^{13}$ cm$^{-2}$ for top and bottom surfaces[1, 3]) indicating that the bulk conduction band must also be populated. The Hall mobility is 1200 cm$^2$/Vs and 300 cm$^2$/Vs for Samples 1 and 2 respectively at $V_g = 0$. The gate-voltage-dependent transport in Sample 1 and 2 is qualitatively similar to that observed by other groups for thicker exfoliated crystals[13-14] and films[12]. While weakly (logarithmically) insulating behavior has been observed in thin $Bi_2Se_3$ films[12, 24], the observation here of transistor-like behavior and a strong (exponentially) insulating state in the thinnest samples is novel, and below we will focus on this behavior in more detail.

Figures 3a and 3b shows the two-probe conductance of Sample 3 as a function of gate voltage $G(V_g)$ at various temperatures, $T$ showing *n*-type field effect behavior. (Similar results were obtained for Sample 4). At high $T$ (245 K – 320 K, Figure 3a) we observe that for positive (negative) $V_g$, the conductance increases (decreases) with decreasing temperature, indicating metallic (insulating) behavior. At lower temperatures (Fig. 3b) the conductance decreases with decreasing temperature at all gate voltages. The maximum field effect mobility is ~10 cm$^2$/Vs at $T = 245$ K.

Figures 3c and 3d show the conductance data from Figs. 2a and b on an Arrhenius plot. At negative gate voltage (Fig. 2c), strongly activated temperature-dependent conductance is observed; straight lines are fits to $G(V_g) = G_0 e^{-E_a/kT}$ where $E_a$ is the activation energy, $k$ is Boltzmann's constant, and $G_0$ a constant prefactor. At positive $V_g$ and lower temperatures (Fig. 2d), activated behavior is also seen with much smaller activation energies.

Figures 4a and 4b show the gate-voltage dependence of the activation energies extracted from Figs. 3c and 3d. For negative gate voltages, the activation energy rises roughly linearly with gate voltage, extrapolating to zero at a threshold $V_g = -10$ V, and rising to 250 meV at $V_g = -90$ V. We interpret the activation energy in this regime as arising due to a barrier to conduction in the bulk; i.e. bulk insulating behavior. (We find the possibility of the activation barrier arising from an insulating contact to a metallic surface state extremely unlikely; first, we observe Ohmic contacts similarly fabricated on slightly thicker $Bi_2Se_3$, and second, we cannot imagine a scenario in which the contact, which lies on top of the sample, could show activation behavior continuously tuned by gate voltage from metallic to insulating.) We assume the activated behavior arises from activation of electrons from the Fermi energy, $E_F$ to conduction band edge, $E_C$; that is $E_a = E_C - E_F$. Then the variation of $E_a$ with $V_g$ reflects the variation of $E_F$: $dE_F/d(eV_g) = -dE_a/d(eV_g)$. The fact that the slope $dE_F/d(eV_g) \ll 1$ indicates movement of Fermi level with back gate through localized impurity states in the band gap. A change in the electrochemical potential of the gate $e\Delta V_g$ is the sum of the electrostatic potential change $e\Delta\varphi$ and the Fermi energy change $\Delta E_F$: $e\Delta V_g = e\Delta\varphi + \Delta E_F = e^2\Delta n/C_g + e^2\Delta n/C_t$ where $\Delta n$ is the change in charge number density, $C_g = 1.15 \times 10^{-8}$ F/cm$^2$ is the oxide capacitance per unit area, and $C_t = e^2 D(E)$ is the quantum capacitance associated with a density of localized states $D(E)$. Then the slope $dE_F/d(eV_g) = C_g/(C_g+C_t)$. From the slope $dE_F/d(eV_g) = 3.3 \times 10^{-3}$, we can estimate $D(E) = 2.1\times 10^{13}$ eV$^{-1}$cm$^{-2}$, and the total charge depleted at $V_g = -90$ V is estimated as $5\times 10^{12}$ cm$^{-2}$ from the bandwidth 250 meV. It is notable that similar behavior was observed in another exfoliated transition-metal chalcogenide, conventional semiconductor MoS$_2$ FETs on SiO$_2$[25], where field-effect mobilities of 10-50 cm$^2$/Vs and a localized state density of 7 x 10$^{12}$ eV$^{-1}$cm$^{-2}$ were measured. Below $T$ = 110K, activated conduction behavior is seen even at positive gate voltage (Fig. 3b) and the activation energy is plotted in Figure 4b. We attribute the very small activation energies in Fig. 3b to a small Schottky barrier between the Pd contacts and the insulating ultrathin Bi$_2$Se$_3$.

The energy barrier in our ultrathin Bi$_2$Se$_3$ FETs is surprisingly large, approaching the bulk energy gap of ~300 meV. We interpret the activation energy as arising from an insulating state in the Bi$_2$Se$_3$, due to coupling of the top and bottom surface states. The magnitude of the energy gap is somewhat larger than the gap of 5 – 50 meV theoretically predicted for 3 QL Bi$_2$Se$_3$[15-17] and the gap of 130 meV observed for 3 QL Bi$_2$Se$_3$ by ARPES experiments[18], though it is comparable to the measured gap for 2 QL Bi$_2$Se$_3$ [18]. This suggests that the significant density of localized states $D(E)$ = 2.1x10$^{13}$ eV$^{-1}$cm$^{-2}$ observed in our experiment may reflect localization of the surface-state-derived bands (which are no longer topologically protected by localization), and conduction may occur in the bulk quantum-well states which should be separated by a gap significantly larger than the bulk gap of 300 meV.

The absence of any *p*-type conduction channel observed up to $V_g$ = -90 V indicates that the actual transport gap may be even larger than 250 meV; in principle one would expect that the high-workfunction Pd contacts would show a smaller barrier for *p*-type injection. The observation of conductance $G$ < 10 nS corresponds to a mean free path for any one-dimensional edge modes < 1 nm; we therefore conclude that it is unlikely that the ultrathin Bi$_2$Se$_3$ is in the quantum spin Hall state.

In conclusion, we have fabricated field-effect transistors from ultrathin Bi$_2$Se$_3$ crystals obtained by mechanical exfoliation. The Bi$_2$Se$_3$ FETs show *n*-type behavior, with a clear insulating OFF state and energy barriers up to 250 meV.  The small subthreshold swing indicates a large density of trap states $D(E) = 2.1 \times 10^{13}$ eV$^{-1}$cm$^{-2}$. The observation of a true insulating state in topological insulator Bi$_2$Se$_3$ is presumed to be due to coupling of the top and bottom surface states, resulting in a conventional two-dimensional insulator.

Acknowledgements.

We acknowledge support from the UMD-NSF-MRSEC grant DMR-05-20471. NPB acknowledges support from CNAM.

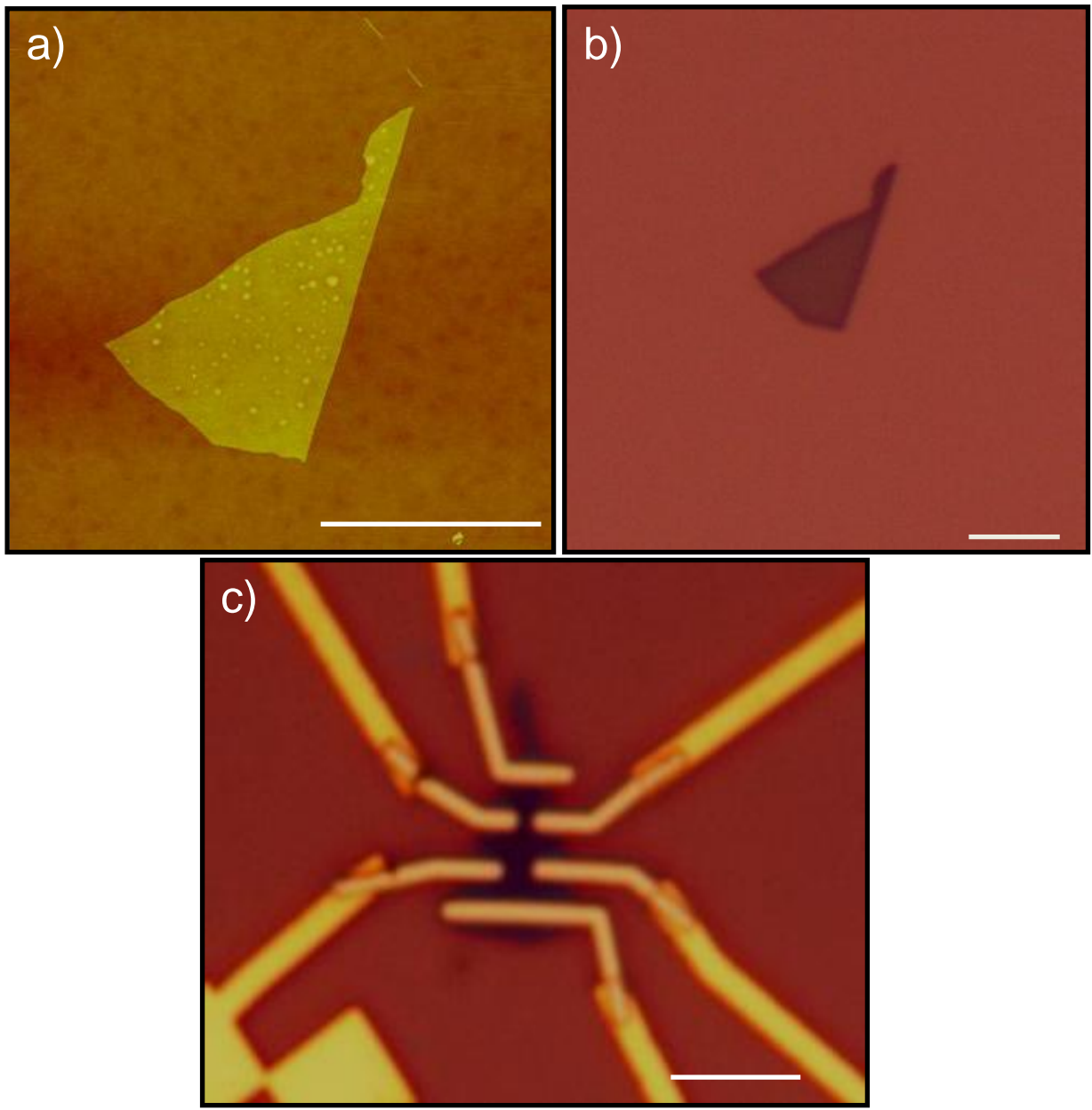

Figure 1. Atomic force micrograph (a) and optical micrographs (b-c) of a 3.5 nm thick exfoliated $Bi_2Se_3$ sample on $SiO_2$/Si substrate. Panel (c) shows the completed device with Pd electrodes contacting the device and larger Cr/Au electrodes leading to bonding pads. Scale bars in (a-c) are 4 microns.

Fig. 2

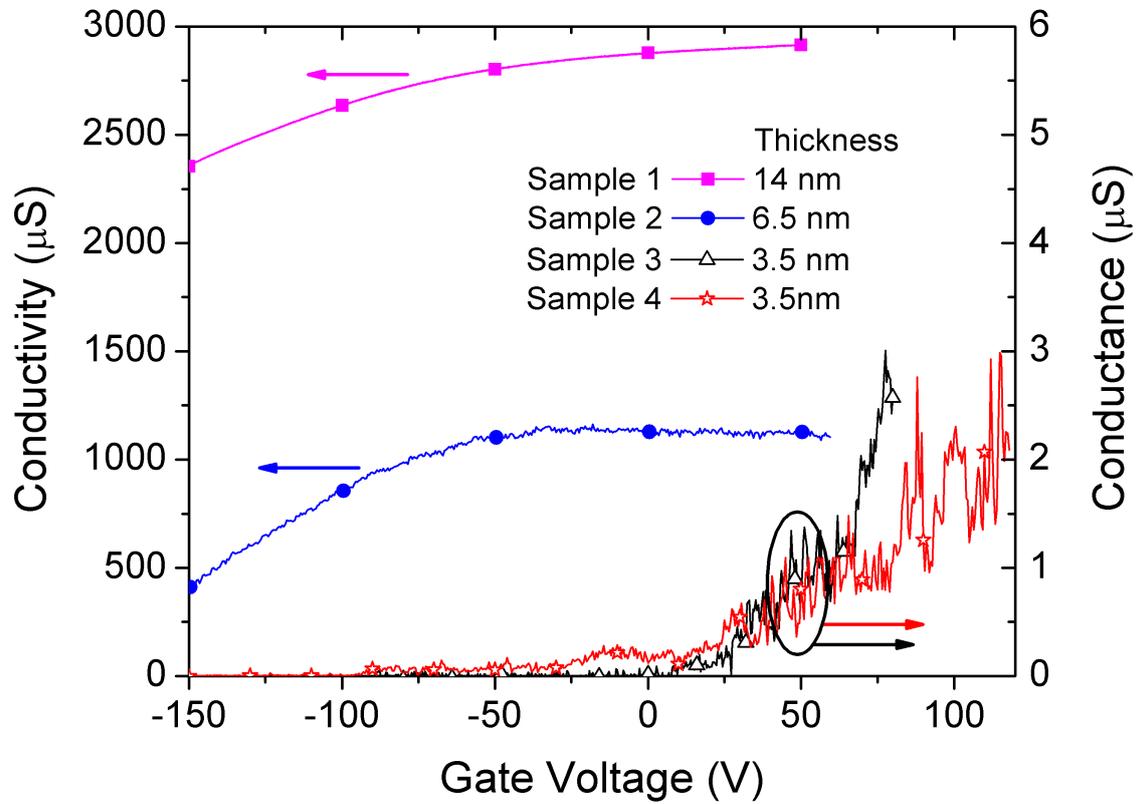

Figure 2. Gate-voltage dependent transport in four exfoliated $Bi_2Se_3$ samples. For thicker Samples 1 and 2, the four-probe conductivity (left axis) as a function of gate voltage is shown. For thinner Sample 3 and 4, the two-probe conductance (right axis) as a function of gate voltage is shown.

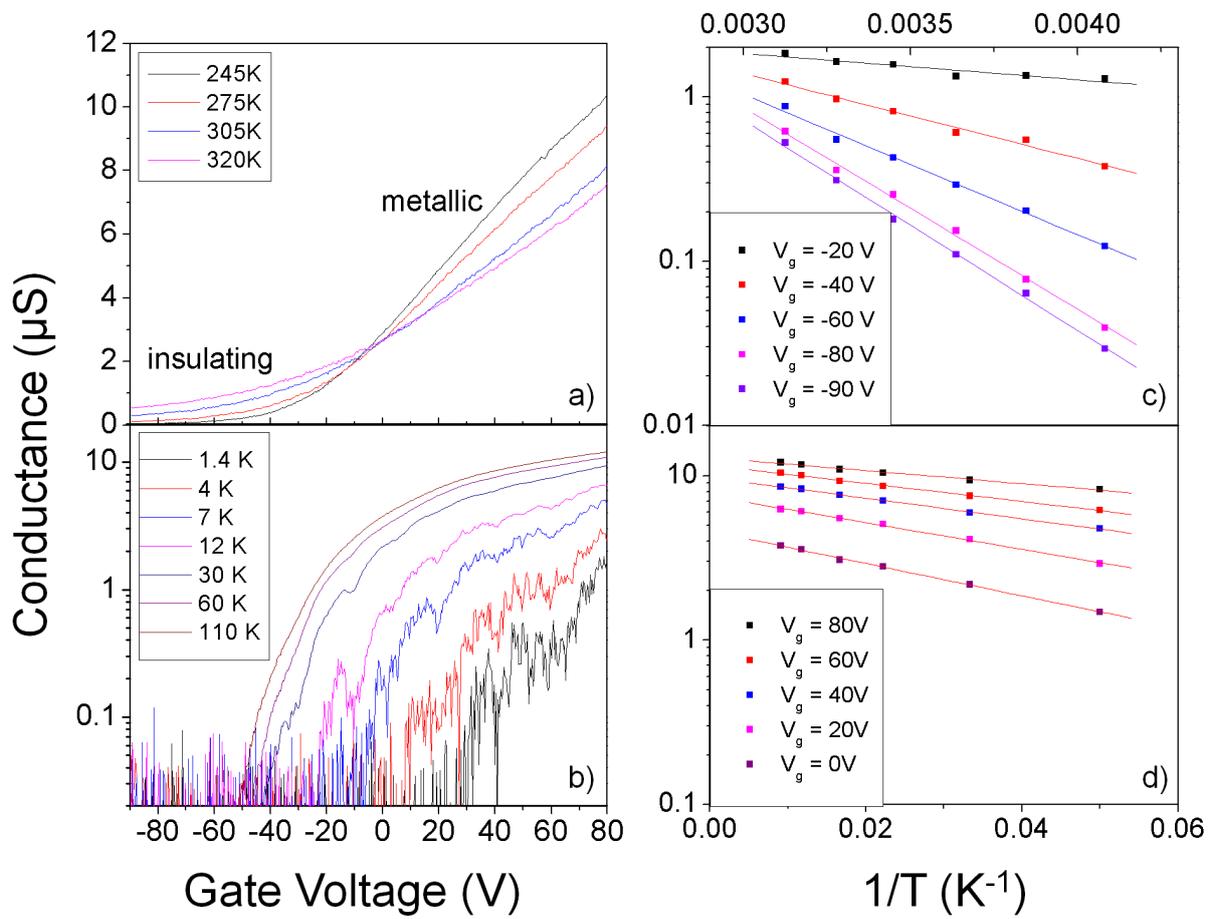

Figure 3. Temperature-dependent conductance of 3.5 nm thick exfoliated $Bi_2Se_3$ on $SiO_2$/Si. (a-b) Conductance of Sample 3 vs. gate voltage at various temperatures. (c-d) Conductance of Sample 3 vs inverse temperature on a semilog scale (Arrhenius plot) showing activated behavior. Lines are linear fits to the data.

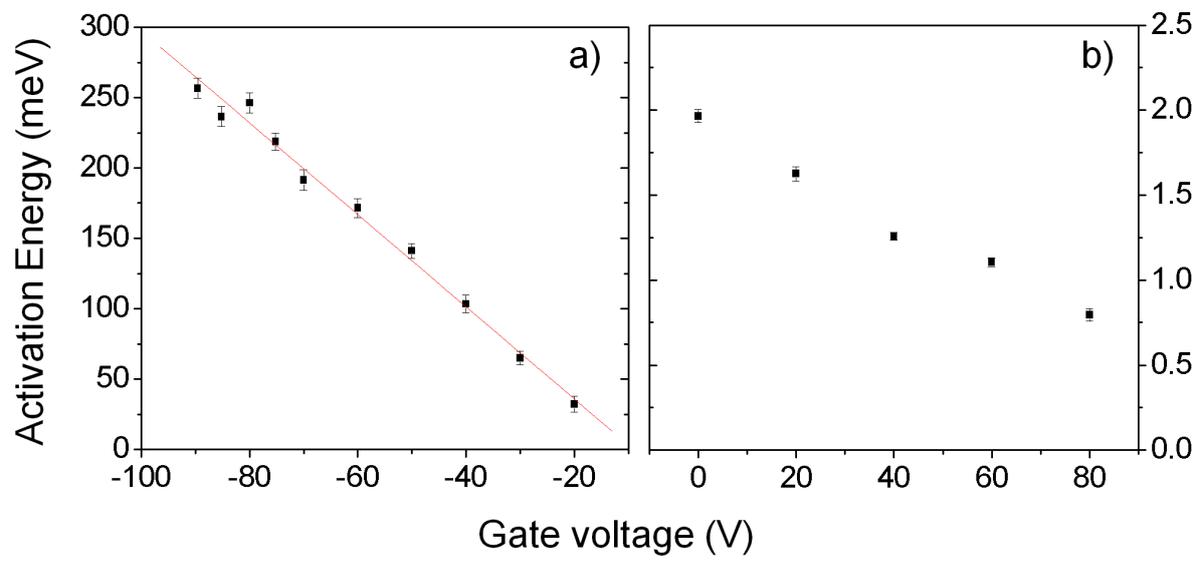

Figure 4. Activation energy as a function of gate voltage determined from fits in Figs. 3c and 3d.